\documentclass[epj]{svjour}
\usepackage{graphicx}
\usepackage{color}
\newcommand{\be}{\begin{equation}}
\newcommand{\ee}{\end{equation}}
\newcommand{\bea}{\begin{eqnarray}}
\newcommand{\eea}{\end{eqnarray}}
\newcommand{\bml}{\begin{mathletters} \baselineskip 10pt}
\newcommand{\eml}{\baselineskip 12pt \end{mathletters}}

\def\lambdabar{\protect\@lambdabar}
\def\@lambdabar{%
\relax \bgroup
\def\@tempa{\hbox{\raise.73\ht0
\hbox to0pt{\kern.2\wd0\vrule width.7\wd0
height.1pt depth.1pt\hss}\box0}}%
\mathchoice{\setbox0\hbox{$\displaystyle\lambda$}\@tempa}%
{\setbox0\hbox{$\textstyle\lambda$}\@tempa}%
{\setbox0\hbox{$\scriptstyle\lambda$}\@tempa}%
{\setbox0\hbox{$\scriptscriptstyle\lambda$}\@tempa}%
\egroup }

\begin{document}
\title{Exploring high-intensity QED at ELI}


\author{Thomas Heinzl\inst{1}\thanks{Speaker}
\and Anton Ilderton\inst{2}
}                     
%
%
\institute{University of Plymouth, School of Mathematics and Statistics, Drake Circus, Plymouth, PL4 8AA, UK
\and School of Mathematics, Hamilton Building, Trinity College, Dublin 2, Ireland
}
\date{Received: date / Revised version: date}
%
\abstract{
We give a non-technical overview of QED effects arising in the presence of ultra-strong electromagnetic fields highlighting the new prospects provided by a realisation of the ELI laser facility.
\PACS{
      {PACS-key}{discribing text of that key}   \and
      {PACS-key}{discribing text of that key}
     } 
} 
\maketitle
\section{Introduction}
\label{intro}

Since the realisation of chirped pulse amplification \cite{Strickland:1985} both power and intensity of lasers have been growing continuously. The present record values of about 1 Petawatt (PW) and $10^{22}$ W/cm$^2$ are expected to be soon superseded, culminating for the foreseeable future at the Extreme Light Infrastructure (ELI) facility with specifications being close to an Exawatt and $10^{26}$ W/cm$^2$. Large intensities correspond to large photon densities so that the corresponding laser beams are very well described as classical (background) field configurations. If, in such a background, we resolve physics at the scale of an electron Compton wavelength we are dealing with QED in the presence of a strong external field or strong-field QED, for short.

\begin{figure}[h]
\resizebox{0.65\columnwidth}{!}{%
\hspace{2cm}  \includegraphics{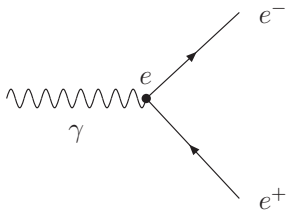}
}
\caption{The elementary vertex in QED couples a photon to an electron and a positron with strength $e$.}
\label{fig:1}       
\end{figure}

Consider the basic QED interaction depicted in Fig.~\ref{fig:1} where a photon couples to an electron and a positron with a coupling strength given by the elementary charge, $e$. The associated pair creation, $\gamma \to e^+ e^-$, cannot happen for real particles as it is forbidden by momentum conservation. A real photon $\gamma$ cannot decay into electron positron pairs.

\begin{figure}[h]
\resizebox{1\columnwidth}{!}{%
\hspace{0cm}  \includegraphics{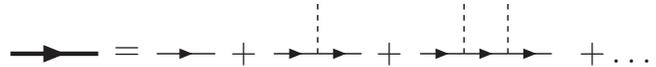}
}
\caption{An electron line dressed by laser photons.}
\label{fig:2}       
\end{figure}

This situation changes, however, in the presence of a laser background which provides an abundance of additional photons. This may be visualised by drawing effective `fat' electron lines which represent dressed particles that continuously absorb and emit laser photons (see Fig.~\ref{fig:2}).

Classically, this corresponds to the quiver motion of electrons once they feel the laser background. All this reinforces the picture that we are discussing a regime interchangeably characterised by strong fields, high intensities and multi-photon effects. These features are all measured by a useful parameter, a dimensionless laser amplitude defined by
\be \label{A0}
  a_0 \equiv \frac{e E \lambdabar_L}{m_e c^2} \; .
\ee
This is the purely classical ratio (no $\hbar$) of the energy gained by an electron traversing the laser field $E$ across a laser wavelength $\lambdabar_L$, and the electron rest energy. For $a_0 > 1$ the electron quiver motion is so rapid that the electron becomes relativistic. We remark that the definition (\ref{A0}) can be generalised to an explicitly gauge and Lorentz invariant expression involving only Minkowski scalars built from probe 4-momenta and field strengths \cite{Heinzl:2008rh}.  In Table~\ref{tab:a0} we give an overview of intensities and $a_0$ values achieved or expected at current and future high-power laser facilities. We have used the rule of thumb, $a_0^2 \simeq 5 \times 10^3 P/\mbox{PW}$,  relating $a_0$ to laser power $P$ in Petawatts \cite{McDonald:1986zz}.

\begin{table}[h]
\renewcommand{\arraystretch}{1.2}
\caption{Current and future laser facilities compared to ELI: intensities $I$ (in W/cm$^2$) and $a_0$ values (XFEL: X-ray free electron laser--DESY, FZD: Forschungszentrum Dresden-Rossendorf, Vulcan: Central Laser Facility--RAL, HiPER: High Power laser Energy Research facility).}
\label{tab:a0}
\begin{tabular}{llllll}
\hline\noalign{\smallskip}
& XFEL  & FZD &  Vulcan  & Vulcan & ELI\\
& (`goal')& (100TW) & (1PW) & (10PW) & HiPER \\
\noalign{\smallskip}\hline\noalign{\smallskip}
$I$  & $10^{27}$ & $10^{20}$ &  $10^{22}$ & $10^{23}$ & $10^{25}$ \\
$a_0$ & $10$  & $20$ & $70$ & $200$ & $5 \times 10^{3}$ \\
\noalign{\smallskip}\hline
\end{tabular}
\end{table}
The small value of $a_0$ for the XFEL is due to its short (X-ray) wave length. Hence, large $a_0$ is exclusively achieved by high-intensity \textit{optical} lasers. In the remainder of this contribution we will give an overview of the most important $a_0$ effects and discuss the prospects for their observation.

\section{Vacuum polarisation effects}
\label{sec:1}

\begin{figure}[h]
\resizebox{1\columnwidth}{!}{%
\hspace{0cm}  \includegraphics{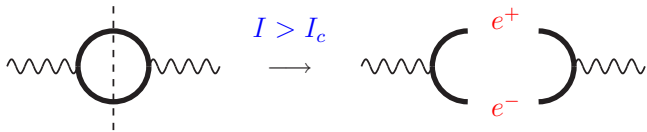}
}
\caption{Dressed vacuum polarisation loop. For sufficiently large intensities the virtual pairs in the loop become real.}
\label{fig:3}       
\end{figure}

Vacuum polarisation is a genuine QED process describing the probability amplitude of a propagating photon `fluctuating' into a virtual electron-positron pair. It has measurable effects such as the Lamb shift and charge screening at short distances (charge renormalisation). Pictorially it is described by the Feynman graph of Fig.~\ref{fig:3} where we have already assumed that the particles in the loop are dressed by the external field as in Fig.~\ref{fig:2}.

\subsection{Pair production}
\label{subsec:PP}

Calculating the graph using Feynman rules we find that, for sufficiently large external fields, it develops an imaginary part when the pair pair creation threshold is exceeded. An analogous statement holds even for vanishing external fields. This is illustrated on the right-hand side of Fig.~\ref{fig:3} which describes the break-up of the loop with virtual pairs becoming real. There is a precise mathematical formulation relating the imaginary part of the loop and the pair creation probability in terms of the optical theorem (a generalisation of Kramers-Kronig relations) which, however, we will not need here. In any case, the real pairs will screen the external electric field and hence lower the overall energy of the system. This is the simple physics of what is sometimes denoted as `vacuum breakdown'. As photons thus `disappear' in favour of pairs the process is called absorptive, and is signalled by the appearance of an imaginary part. The relevant scale for this is set by the critical electric field \cite{Sauter:1931zz},
\be \label{EC}
  E_c = \frac{m_e^2  c^3}{e  \hbar} \simeq 1.3 \times 10^{18} \; \, \mbox{V/m} \; .
\ee
In such a field an electron would gain an energy equal to its rest energy upon travelling across a Compton wavelength. The presence of both $c$ and $\hbar$ in (\ref{EC}) calls for a unification of special relativity and quantum mechanics which is precisely achieved by quantum field theory, in the present context QED. The critical intensity corresponding to (\ref{EC}) is $I_c \simeq 4 \times 10^{29}$ W/cm$^2$ which is, presumably, beyond reach within the next decade or so.

\begin{figure}[h]
\resizebox{0.7\columnwidth}{!}{%
\hspace{2cm}  \includegraphics{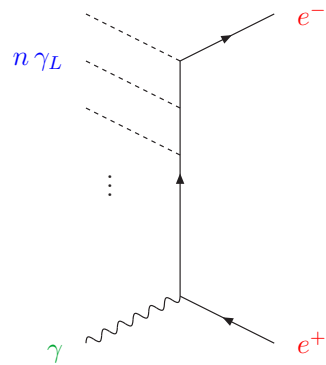}
}
\caption{Perturbative multi-photon contribution to pair production.}
\label{fig:4}       
\end{figure}

The physics of vacuum polarisation effects is governed by two dimensionless parameters which we call $\nu$ and $\epsilon$, defined as
\be \label{NUEPSILON}
  \nu \equiv \frac{\hbar\omega}{m_e c^2} \; , \quad \epsilon \equiv \frac{E}{E_c} = \frac{\hbar \omega_L}{m_e c^2} \, a_0 \simeq 10^{-6} a_0 \; .
\ee
$\nu$ measures the energy of the propagating (and possibly `decaying') photon in units of the electron rest energy while $\epsilon$ is the ratio of the electric field and Sauter's critical field (\ref{EC}). The latter can be traded for $a_0$ with a small prefactor measuring the energy of laser light (of order 1~eV) in terms of $m_e c^2$. Pair creation may then be due to large $\nu > 1$ (energy above threshold), large fields, $\epsilon > 1$ and $a_0 > 10^6$, or a combination of both. There has been a single experiment that has detected laser induced pair production, namely SLAC E-144 \cite{Burke:1997ew,Bamber:1999zt}. This was done using a Terawatt laser and the SLAC beam and hence was a high energy, low intensity experiment with $\nu \simeq 6 \times 10^4$ and $a_0 \simeq 0.4$. A typical higher-order process contributing to pair creation is depicted in Fig.~\ref{fig:4} which resolves the dressed electron line in terms of $n$ laser photons $\gamma_L$.

We thus have $\gamma + n \gamma_L \to e^+ e^-$, which represents a multi-photon generalisation of Breit-Wheeler pair production \cite{Breit:1934} (see also \cite{Reiss:1962}). The high energy probe photon $\gamma$ was produced via nonlinear Compton backscattering (see Sect.~\ref{sec:2}) off the SLAC beam. To overcome the threshold a minimal photon number of $n_0 = 5$ was required and confirmed by checking the behaviour of the production rate, $R \sim a_0^{2 n_0}$. This power law behaviour is \textit{perturbative} in nature as is to be expected for $a_0 \ll 1$ \cite{Dunne:2004nc,Dunne:2008}. No experiment has ever looked for pair creation at high intensities (large $a_0$). In this case, there are still two options: (i) energy above threshold ($\nu > 1$) and subcritical fields ($a_0 \gg 1$, but $\epsilon < 1$) or (ii) energy below threshold ($\nu < 1$) and supercritical fields ($a_0 \gg 1$, $\epsilon > 1$). Scenario (i) is easier to realize and will be briefly discussed in the next subsection. Case (ii) is the most spectacular and corresponds to Schwinger pair production \cite{Schwinger:1951nm} by spontaneous vacuum decay. This is a completely nonperturbative process with an exponentially suppressed pair creation rate, $R \sim \exp(-\pi/\epsilon)$, typical for tunneling processes. Note that it cannot be expanded in $\epsilon$ or $a_0$. It is covered thoroughly by G.~Dunne in these proceedings \cite{Dunne:2008}. We will move on to the analysis of the real part of the polarisation loop of Fig.~\ref{fig:3}.

\subsection{Vacuum birefringence}
\label{subsec:VB}

\begin{figure}[h]
\resizebox{1\columnwidth}{!}{%
\hspace{0cm}  \includegraphics{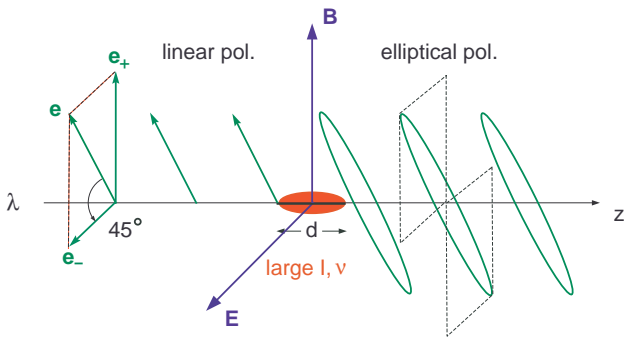}
}
\caption{Schematic experimental setup for a vacuum birefringence measurement.}
\label{fig:5}       
\end{figure}

The real part of the vacuum polarisation diagram describes photon propagation modified by virtual pairs. As such it corresponds to a \textit{dispersive} process. Strong background fields actually modify the dispersion relation for photons in a peculiar way. As first discussed by Toll in his unpublished thesis \cite{Toll:1952} the external field with its preferred direction induces `vacuum birefringence', i.e.\ nontrivial refractive indices which differ for different polarisation directions of the probe photons. To lowest order in $\nu$ and $\epsilon$ the two principal indices are \cite{Toll:1952,Heinzl:2006pn}
\be \label{NPM}
   n_\pm = 1 + \frac{\alpha\epsilon^2}{45 \pi}  \Big\{ 11 \pm 3   + O( \epsilon^2  \nu^2) \Big\} \Big\{ 1 + O(\alpha\epsilon^2) \Big\}\; ,
\ee
where $\alpha = e^2/4\pi\hbar c \simeq 1/137$ denotes the fine structure constant as usual. As $\alpha\epsilon^2 \ll 1$ the deviation from unity is basically a function of the product $\epsilon\nu$ (first curly bracket) and, even for ELI intensities ($\epsilon \simeq 10^{-2}$), extremely small, of the order of $10^{-8}$. Nevertheless, it is not hopeless to attempt a measurement \cite{Heinzl:2006xc}. The setup for a vacuum birefringence experiment is shown in Fig.~\ref{fig:5}.

A linearly polarised probe beam collides head-on with a laser of ultra-high intensity $I$ concentrated within a focus size $d$ (taken to be the Rayleigh length). The relative phase retardation due to birefringence induces a small but nonvanishing ellipticity once the beam has passed the high-field region. For $\epsilon, \nu \ll 1$ the signal is given by the expression
\be \label{ELLIPTICITY}
  \delta^2 = 3.2 \times 10^5 \left( \frac{d}{\mathrm{\mu m}}  \epsilon^2  \nu \right)^2 \; ,
\ee
and hence power law suppressed. For an X-ray probe (i.e.\ $\nu \simeq 10^{-2}$), a Rayleigh length $d \simeq 10 $ $\mu$m and ELI intensities one finds an ellipticity signal $\delta^2$ of the order of $10^{-5}$. This is at the lower end of what can nowadays be measured using X-ray polarimetry \cite{Alp:2000}. Assuming a larger Rayleigh length one can gain an order of magnitude so that the signal could safely be detected with present day techniques. The result (\ref{ELLIPTICITY}) is valid for small $\nu$ and $\epsilon$ (low energy and intensity). While the field strength $\epsilon$ cannot be increased easily we may, however, consider large probe frequency, $\nu > 1$, hence scenario (i) of the preceding subsection. For this we need polarised high-energy photons. The standard method to produce these is via Compton backscattering off an electron beam \cite{Milburn:1962jv,Ballam:1969jr}. This was also the method of choice for the SLAC E-144 experiment \cite{Bula:1996st}. To achieve high photon energies (i.e.\ a Compton blue shift, hence `inverse' Compton scattering) it is best to use another laser with $a_0 < 1$ (see Sect.~\ref{sec:2}). The maximum frequency is then given by the Compton edge $\omega \simeq 4 \gamma^2 \omega_0$ with $\hbar \omega_0 \simeq 1$ eV for optical lasers. To exceed the pair creation threshold hence requires electrons of a minimal energy of $E_e \simeq 250$ MeV. This is nowadays routinely achieved using laser wake field acceleration (or modern variants thereof).
\begin{figure}[h]
\resizebox{0.8\columnwidth}{!}{%
\hspace{2cm}  \includegraphics[angle=270]{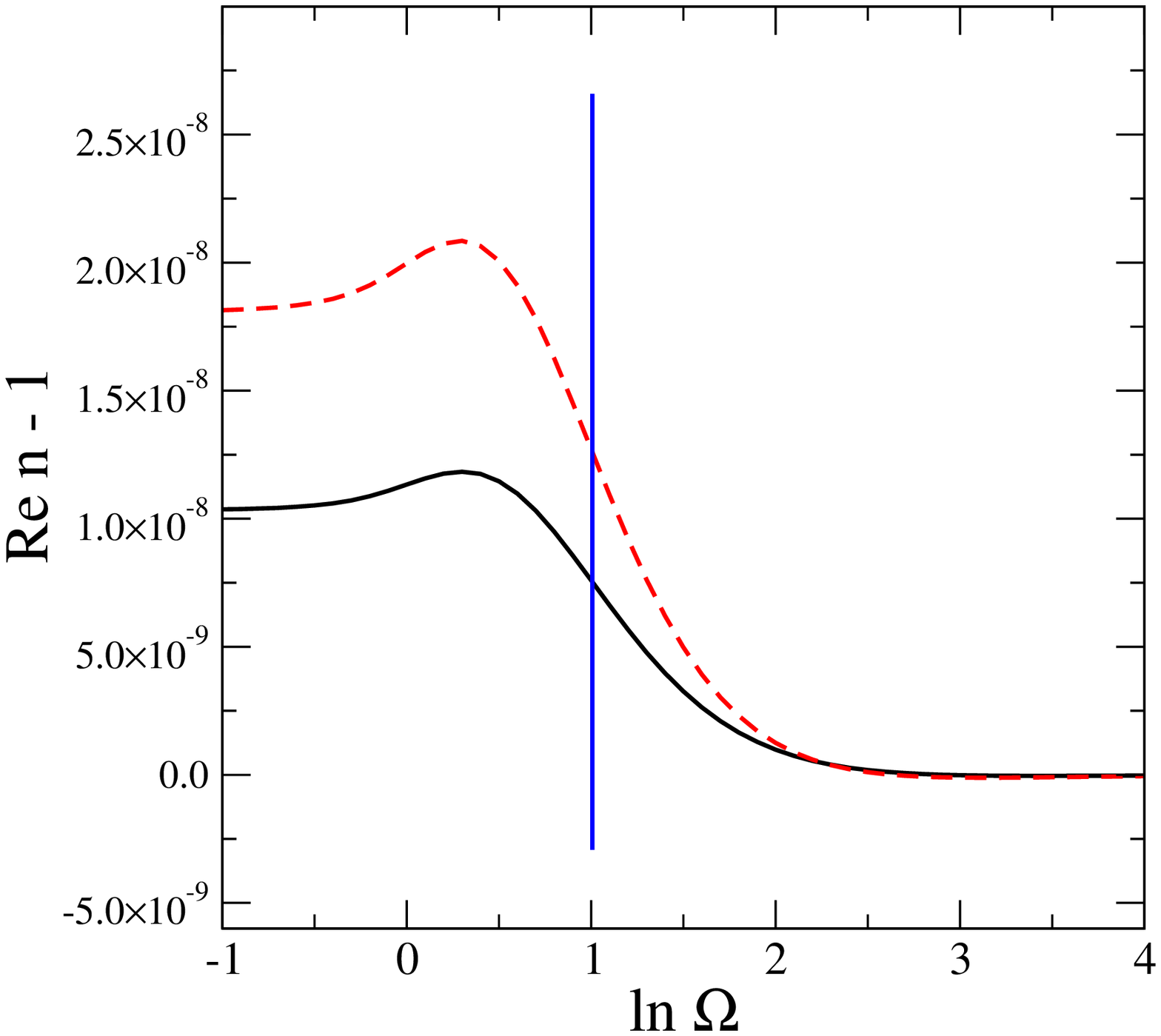}}
\vspace{.5cm}
\resizebox{0.8\columnwidth}{!}{%
\hspace{2cm}  \includegraphics[angle=270]{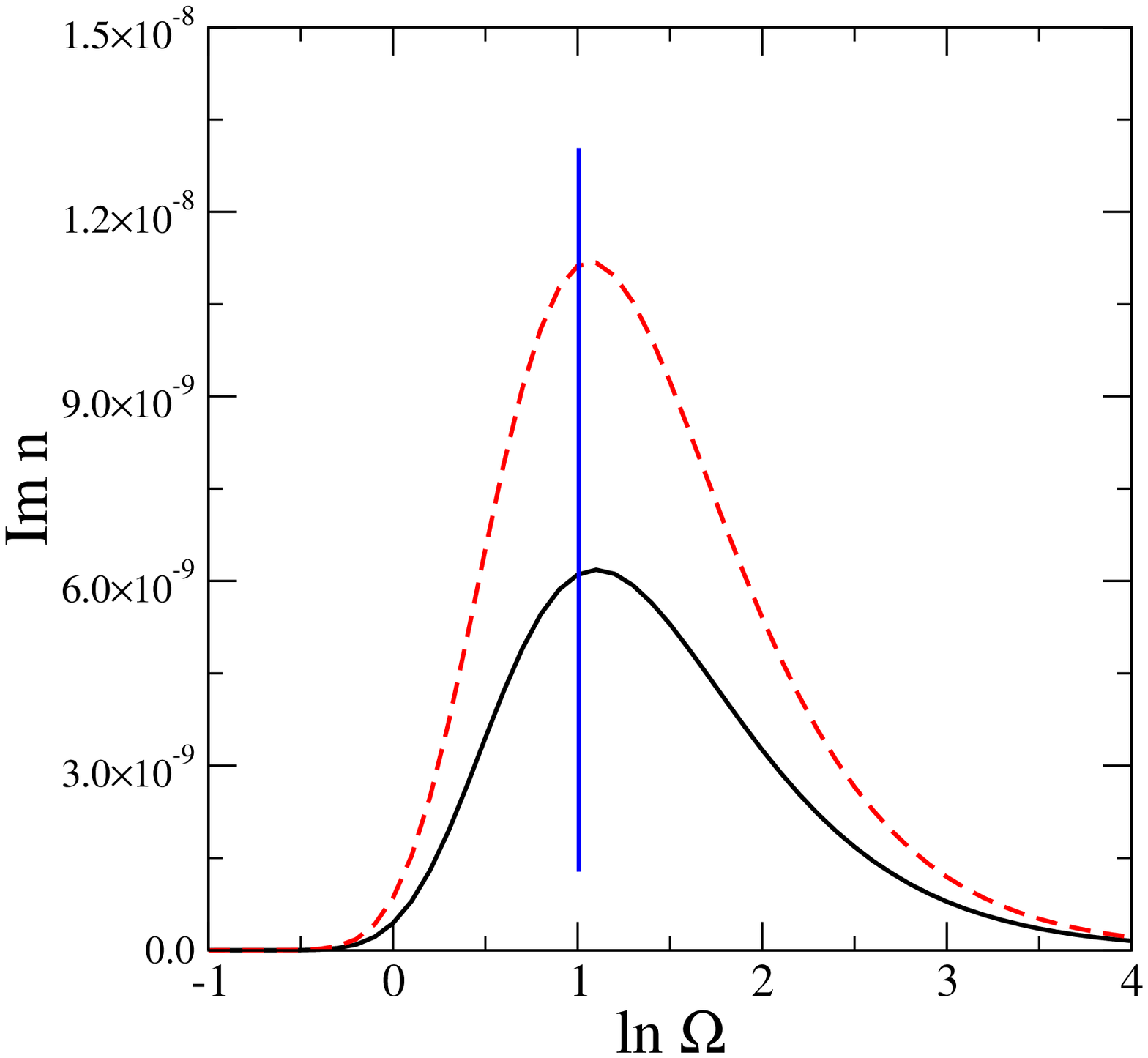}
}
\caption{Real and imaginary parts of the QED refractive indices as a function of $\ln \Omega \equiv \ln \epsilon\nu$. Dashed line: $n_+$, full line: $n_-$, vertical line: $\ln \Omega = 1$, achieved for photons backscattered off 3 GeV electrons.}
\label{fig:6}       
\end{figure}

For definiteness, let us assume an electron energy of $E_e = 3$ GeV. For ELI intensities ($\epsilon \simeq 10^{-2}$) we then have $\epsilon\nu \simeq 2.7 > 1$ and so the expansion (\ref{NPM}) no longer makes sense. Fortunately, for $\epsilon < 1$, the indices are known for all values of $\epsilon\nu$, as first worked out by Toll \cite{Toll:1952} (see also \cite{Shore:2007um}). Both their real and imaginary parts are depicted in Fig.~\ref{fig:6}. Note that, for $\epsilon\nu \simeq 2.7$, we are able to access and test a highly nontrivial region where the real parts of the indices show a negative slope, hence \textit{anomalous} dispersion. By Kramers-Kronig this is associated with a nonvanishing imaginary part, an alternative signal for pair production. So, with ELI it seems feasible to study the frequency dependence of the QED refractive indices for the first time.

\section{Nonlinear Compton scattering}
\label{sec:2}

\begin{figure}[h]
\resizebox{1\columnwidth}{!}{%
\hspace{0cm}  \includegraphics{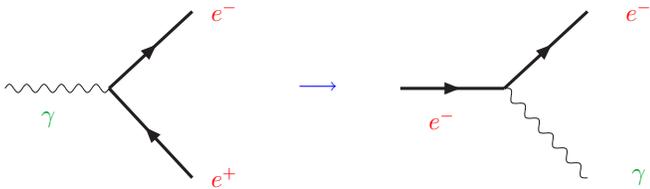}
}
\caption{From multi-photon pair creation to nonlinear Compton scattering via crossing.}
\label{fig:7}       
\end{figure}

All effects and processes discussed so far are typically power-law or exponentially suppressed (in the perturbative or nonperturbative regime, respectively). In addition, perturbative pair production can only proceed once the energy threshold is overcome. As a result, both dispersive and absorptive vacuum polarisation phenomena are small and, in general, difficult to observe. One may therefore ask whether there are any intensity dependent effects that are more accessible to experimental observation, even with currently existing laser specifications. It turns out that the answer to this question is yes -- a readily observable such process is nonlinear Compton scattering consisting of processes of the type
\be
  e + n \gamma_L \to e' + \gamma \; ,
\ee
with $n$ counting the number of laser photons involved. Formally, this process is obtained from multi-photon Breit-Wheeler pair production via crossing, i.e.\ exchanging a photon and a lepton line, see Fig.~\ref{fig:7}. Note that the laser photons are `hidden' in the dressed electron lines.

\begin{figure}[h]
\resizebox{0.95\columnwidth}{!}{%
\hspace{0cm}  \includegraphics{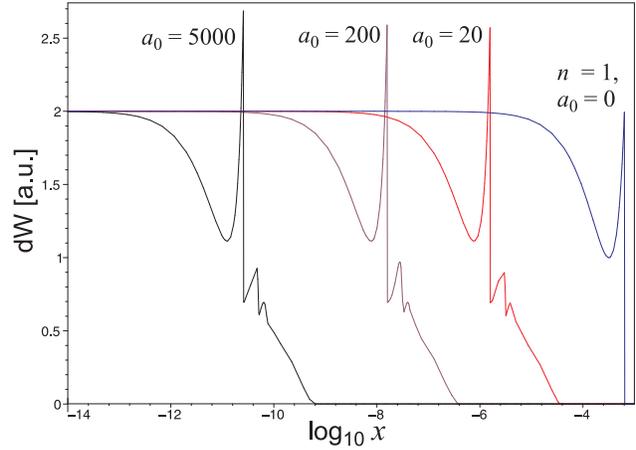}
}
\caption{Photon spectra for (nonlinear) Compton scattering. Note the redshift of the kinematic edges as compared to linear Compton scattering ($n=1$, $a_0 = 0$).}
\label{fig:8}       
\end{figure}

As the Feynman graph suggests a high-intensity laser is scattered off an electron beam which, in an all-optical setup, may be generated via laser wake field acceleration. The main intensity effect (predicted a long time ago \cite{Sengupta:1949}) is a mass shift of the electron depending on $a_0$ according to $m_*^2 = m^2 (1 + a_0^2)$. This results in a red shift of the kinematic edge for ordinary (linear) Compton scattering ($n=1$, $a_0 = 0$) by a factor of $1/a_0^2$ for large relativistic gamma factors, $\gamma \gg 1$,
\be \label{REDSHIFT}
  4 \gamma^2 \omega_0 \to 4 \gamma^2 \omega_0/a_0^2 \; .
\ee
Here, $\hbar \omega_0 \simeq 1$ eV is the energy of the incident laser photons. For a few relevant values of $a_0$ the resulting photon spectra are depicted in Fig.~\ref{fig:8} as a function of the Lorentz invariant $x \equiv k \cdot k'/k \cdot p'$ with $k$, $k'$ and $p'$ denoting the 4-momenta of incident photons, scattered photon and scattered electron, respectively.

The spectra seem to scale with intensity parameter $a_0$ and roughly follow the line shape of linear Compton scattering with a significant peak corresponding to the kinematical edge for one-photon scattering ($n=1$). To the right of this peak, however, there are side maxima corresponding to higher harmonics which decrease towards higher $n$ and $x$.

\begin{figure*}[t]
\begin{center}
\includegraphics[scale=0.8]{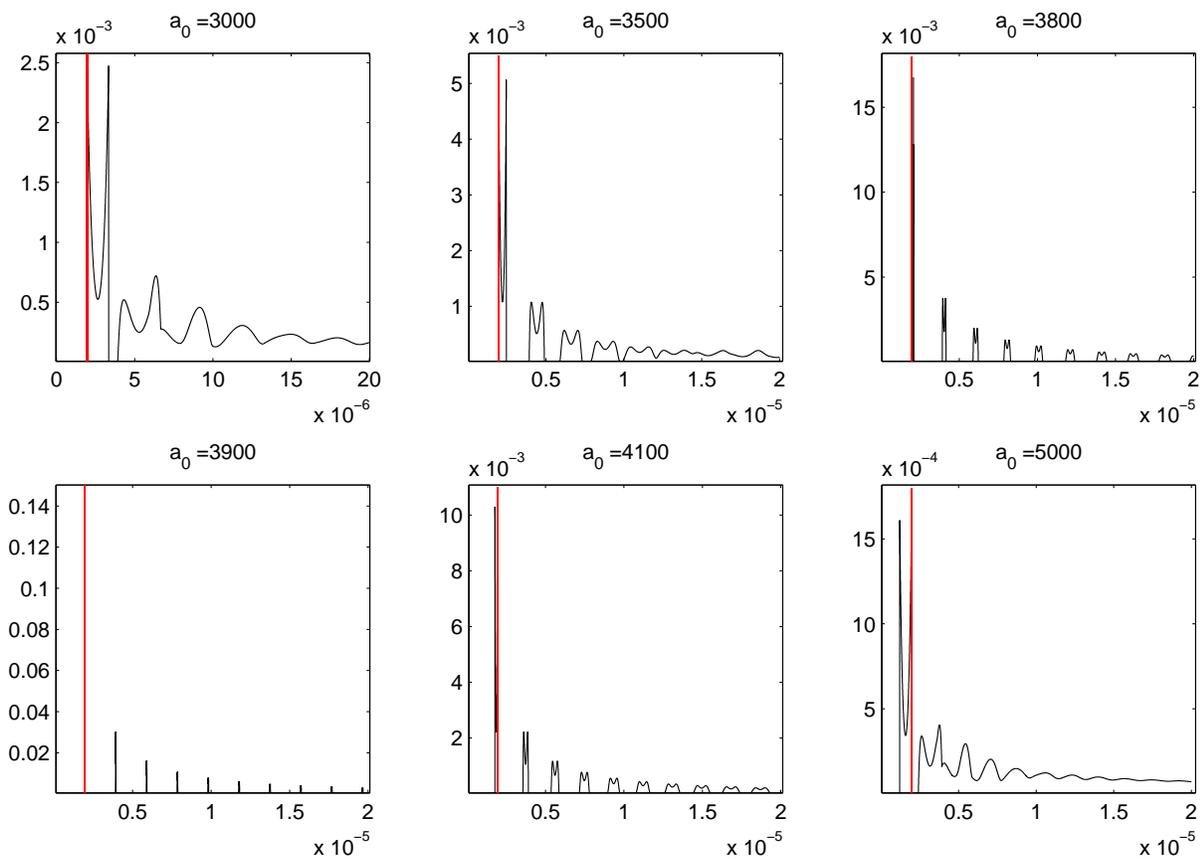}
\end{center}
\caption{Photon spectra for (nonlinear) Compton scattering (off 1 GeV electrons) as a function of dimensionless scattered frequency, $\nu'$, for different values of $a_0$. The vertical (red) lines denote the initial photon frequency, $\nu \equiv \hbar\omega_0/mc^2 = 2 \times 10^{-6}$, corresponding to $\hbar\omega_0 = 1\, $eV. For a critical value of $a_0 \simeq 3900$, $\nu_1' = \nu$, and one obtains essentially a line spectrum for low harmonics  (first `snapshot' in second row).}
\label{fig:9}       
\end{figure*}

Any measurement of the spectra will be done in the lab (frame). In this case one may record photon yields as a function of scattered frequency, $\omega'$ which, for $a_0^2 \ll 4 \gamma^2$, results in plots similar to Fig.~\ref{fig:8}. At a critical value, $a_0^2 \simeq 4 \gamma^2 \gg 1$, however, the $n$=1 red-shift (\ref{REDSHIFT}) becomes so large that incident and scattered photon energies coincide, $\omega_1' = \omega_0$. This demarcates the boundary between an overall blue-shift for one-photon scattering ($\omega_1' > \omega_0$), often called \textit{inverse} Compton scattering, and an overall red-shift ($\omega_1' < \omega_0$) as in Compton's original experiment (with the electron initially at rest). As long as $\gamma \gg n\nu $, where $\nu \equiv \hbar\omega_0/mc^2$ as before, the critical $a_0$ is $n$-independent. Hence, also the scattered frequencies of the higher harmonics ($1 < n \ll \gamma/\nu$) are, to a very good approximation, fixed by the kinematics, $\omega_n' \simeq n \omega_0$. Accordingly, the spectral ranges for the low harmonics shrink to (almost) zero, and one essentially obtains an equidistant line spectrum for them. This is nicely borne out in Fig.~\ref{fig:9} which shows the photon spectra as functions of $\nu' \equiv \hbar\omega'/mc^2$ for different values of $a_0$ and a  incident electron energy of 1 GeV. The corresponding rough estimate for the `critical' boundary value is $a_0 \simeq 2 \gamma \simeq 4 \times 10^3$ which is in good agreement with the numerically precise `thumbnail' pictures of Fig.~\ref{fig:9}.

\section{Summary}

In this brief overview we have discussed a few strong-field QED phenomena that are possibly relevant for the planned ELI facility. The (arguably) most exciting of these, Schwinger pair production, is also the most difficult to observe as even at ELI the fields will be about three orders of magnitude below Sauter's critical value. Surprises are, however, possible, for instance if pre-exponential factors become large \cite{Bulanov:2004de} or deviations from the crossed field limit become significant \cite{Schutzhold:2008pz}. Using the experimental setting of the SLAC E-144 experiment, ideally with the electron linac replaced by a laser (wake field) accelerator, one may use ELI to investigate vacuum birefringence at intermediate energies, with Compton backscattered photon energies above pair creation threshold. This would amount to being perturbative in intensity but nonperturbative in photon frequency. Thus, one may study the frequency dependence of the QED refractive indices with anomalous dispersion signalling absorption due to pair production. Such an experiment would involve nonlinear Thomson or Compton scattering, hence a careful study of this process is required. This programme may actually be launched right away as there are no thresholds to be overcome. As the scattered photon spectra depend most sensitively on intensity they may provide a useful `standard candle' to determine the intensity parameter $a_0$, in particular at ultra-high (ELI) intensities. It seems fair to conclude that we are facing exciting times---for both theory and experiment.

\medskip

\textbf{\textsf{Acknowledgements:}} TH thanks D.\ Habs for the invitation to the ELI workshop and school and the organising team for their outstanding efforts. He is indebted to H.\ Reiss for sending a copy of Toll's thesis. Many thanks to K.\ Langfeld and C.\ Harvey for producing Fig.s~6 and 9, respectively.

%


\end{document}